\documentclass[epj]{svjour}
\usepackage{array}
\usepackage{amsmath}
\usepackage{graphicx}
\usepackage{amssymb}
\usepackage[colorlinks]{hyperref}
\usepackage[normalem]{ulem}

\providecommand{\apj}{Astrophys.\ J.}

\providecommand{\apjs}{Astrophys.\ J.\ Suppl.}

\providecommand{\physrep}{Phys.\ Rep.}
\providecommand{\nat}{Nature}

\begin{document}

\bibliographystyle{ieeetr}

\title{Constraining the supersaturation density equation of state from core-collapse supernova simulations?}
\subtitle{Excluded volume extension of the baryons}
\author{Tobias Fischer
}                     
\offprints{}          
\institute{University of Wroc{\l}aw, Pl. M. Borna 9, 50-204 Wroclaw, Poland}
\date{Received: date / Revised version: date}
\abstract{
In this article the role of the supersaturation density equation of state (EOS) is explored in simulations of failed core-collapse supernova explosions. Therefore the nuclear EOS is extended via a one-parameter excluded volume description for baryons, taking into account their finite and increasing volume with increasing density in excess of saturation density. Parameters are selected such that the resulting supernova EOS represent extreme cases, with high pressure variations at supersaturation density which feature extreme stiff and soft EOS variants of the reference case, i.e. without excluded volume corrections. Unlike in the interior of neutron stars with central densities in excess of several times saturation density, central densities of core-collapse supernovae reach only slightly above saturation density. Hence, the impact of the supersaturation density EOS on the supernova dynamics as well as the neutrino signal is found to be negligible. It is mainly determined from the low- and intermediate-density domain, which is left unmodified within this generalized excluded volume approach.
\PACS{
{26.50.+x}{Nuclear physics aspects of novae, supernovae, and other explosive environments}
\and
{26.60.Kp}{Equations of state of neutron-star matter}
\and
{97.60.Bw}{Supernovae}
     } 
} 
\maketitle

\section{Introduction}
\label{intro}

A neutron star is born in the violent event of a core-collapse supernova explosion of a star more massive than about 9~M$_\odot$. It is associated with the revival of the stalled bounce shock, which forms when the initially imploding stellar core bounces back at supersaturation density, leading to the ejection of the stellar mantle~(for recent reviews c.f. Refs.~\cite{Janka:2007,Janka:2012}). Several scenarios for the shock revival have been discovered~\cite{LeBlanc:1970kg,Burrows:2005dv,Bethe:1985ux,Sagert:2008ka}. Among them the most promising one is due to neutrino heating. However, neutrino-driven explosions generally require multi-dimensional simulations where in the presence of convection and potentially hydrodynamics instabilities  the neutrino heating efficiency increases~\cite{Marek:2009,Mueller:2012b,Bruenn:2013} in comparison to the spherically symmetric case. The exception is the low-mass range, in particular the O-Ne core progenitor of 8.8~M$_\odot$~\cite{Nomoto:1984,Kitaura:2006,Fischer:2009af} and the zero-metallicity Fe-core progenitor of 9.6~M$_\odot$~\cite{Melson:2015}. This is a hot and active subject of research~\cite{Jones:2013}.  More massive progenitors have largely extended silicon layers above the stellar core which make it more difficult to revive the standing bounce shock. The impact of the core-collapse supernova progenitor structure on the supernova dynamics is generally not answered yet~\cite{Couch:2013a,Mueller:2015}.

The role of the EOS in core-collapse supernova simulations was explored even in the multi-dimensional framework~\cite{Marek:2008qi,Suwa:2013,Couch:2013b}, as well as in failed core-collapse supernova explosions in spherical symmetry~\cite{Sumiyoshi:2006id,Fischer:2009,OConnor:2011,Steiner:2013}. Recently, the role of the nuclear symmetry energy has been reviewed~\cite{Fischer:2014}. Particular focus has been devoted to study the role of the high-density behavior exploring therefore the EOS of Ref.~\cite{Lattimer:1991nc} which is available for three different (in)compressibility modulus~\cite{Swesty:1994,Thompson:2003}. Unlike in multi-dimensional simulations where small variations as initial perturbations can grow to large scale effects, e.g., due to the turbulent cascade, in spherically symmetric simulations the role of the high-density EOS was never reported to be significant. Despite large variations of the nuclear matter properties differences of the neutrino signal and the general supernova evolution were on the order of a few \%.

However, a systematic study of the supersaturation density EOS and the impact in core-collapse supernova simulations has not been conduced, which is the aim of this article. Previous studies explored selected EOS which usually differ in all nuclear matter properties. It was therefore not possible to exclusively relate results from supernova simulation to the high-density EOS. Here I follow a different approach by modifying only the supersaturation density EOS of a well selected nuclear relativistic mean-field (RMF) model~\cite{Typel:2005,Hempel:2009mc} that has been widely explored in the core-collapse supernova community~(c.f. Ref.~\cite{Hempel:2012}). Therefore, the novel excluded volume approach introduced in Ref.~\cite{Typel:2015} is employed here for densities in excess of nuclear saturation density ($\rho_0$). It modifies the available volume of the nucleon gas which effectively adjusts the baryon EOS. This setup will allow for a direct identification of the high-density EOS impact on the supernova dynamics as well as the neutrino signal. A preliminary version of this novel excluded volume approach has been discussed recently~\cite{Benic:2015}.

In this study the general relativistic neutrino radiation-hydrodynamics model AGILE-BOLTZRTRAN is used for the supernova simulations. It is based on three-flavor Boltzmann neutrino transport~\cite{Liebendoerfer:2004}, being ideal to study the early post-bounce phase prior to the possible onset of shock revival. The restriction to spherical symmetry is not problematic here since the focus of this study is the supersaturation density EOS where multi-dimensional phenomena can be neglected. Moreover, in this parametric study I will analyze results relative to the reference case -- claims about the magnitude of potential observables render irrelevant.

The manuscript is organized as follows. In sec.~\ref{SNmodel} I will briefly review the spherically symmetric supernova model, including weak reactions and EOS. The subsequent sec.~\ref{ev} briefly introduces the excluded-volume mechanism of \cite{Typel:2015} that is applied to modify the supersaturation density EOS. These are included in simulations of failed core-collapse supernova simulations which are then analyzed in sec.~\ref{SNsimulations}. The paper closes with a summary in sec.~\ref{summary}.

\section{Supernova model}
\label{SNmodel}

The core-collapse supernova model, AGILE-BOLTZTRAN, is based on spherically symmetric and general relativistic neutrino radiation hydrodynamics. It includes angle- and energy-dependent three flavor Boltzmann neutrino transport~\cite{Liebendoerfer:2001a,Liebendoerfer:2001b,Liebendoerfer:2002}. The implicit method for solving the hydrodynamics equations and the Boltzmann transport equation on an adaptive Lagrangian mass grid agreed qualitatively well with other methods, e.g., with the multi-group flux limited diffusion approximation~\cite{Liebendoerfer:2004} and the variable Eddington factor technique \cite{Liebendoerfer:2005a}.
\begin{table}[htp]
\centering
\caption{Neutrino reactions considered, including references.}
\begin{tabular}{ccc}
\hline
\hline
& Weak process & References \\
\hline
1 & $e^- + p \rightleftarrows n + \nu_e$ & \cite{Reddy:1998,Horowitz:2001xf} \\ 
2 & $e^+ + n \rightleftarrows p + \bar\nu_e$ & \cite{Reddy:1998,Horowitz:2001xf} \\
3 & $\nu_e + (A,Z-1) \rightleftarrows (A,Z) + e^-$ & \cite{Juodagalvis:2010} \\
4 & $\nu + N \rightleftarrows \nu' + N$ & \cite{Bruenn:1985en,Mezzacappa:1993gm,Horowitz:2001xf} \\
5 & $\nu + (A,Z) \rightleftarrows \nu' + (A,Z)$ & \cite{Bruenn:1985en,Mezzacappa:1993gm} \\
6 & $\nu + e^\pm \rightleftarrows \nu' + e^\pm$ & \cite{Bruenn:1985en},
\cite{Mezzacappa:1993gx} \\
7 & $e^- + e^+ \rightleftarrows  \nu + \bar{\nu}$ & \cite{Bruenn:1985en} \\
8 & $N + N \rightleftarrows  \nu + \bar{\nu} + N + N $ & \cite{Hannestad:1997gc} \\
9 & $\nu_e + \bar\nu_e \rightleftarrows  \nu_{\mu/\tau} + \bar\nu_{\mu/\tau}$ & \cite{Buras:2002wt,Fischer:2009} \\
10 & $\nu + \bar\nu + (A,Z) \rightleftarrows (A,Z)^*$ & \cite{Fuller:1991,Fischer:2013} \\
\hline
\end{tabular}
\\
$\nu=\{\nu_e,\bar{\nu}_e,\nu_{\mu/\tau},\bar{\nu}_{\mu/\tau}\}$ and $N=\{n,p\}$
\label{tab:nu-reactions}
\end{table}

\subsection{Weak interactions}
\label{weak}

Table~\ref{tab:nu-reactions} lists the set of weak processes considered, including references. Note that weak reactions with heavy nuclei, $e^-$-captures and (de)excitations, are relevant only during the core-collapse phase. Shortly before and after core bounce heavy nuclei are not abundant anymore, in particular in excess of $\rho_0$ and at high temperatures on the order of 10~MeV. At such conditions the nuclear composition is dominated by (partly) dissociated matter with free neutrons and protons as well as light nuclear clusters. During the (early) post-bounce evolution weak reactions with free nucleons are the most important ones. Scattering on neutrons has the largest opacity in the elastic channel while charged-current absorptions on neutrons for $\nu_e$ and protons for $\bar\nu_e$ has largest opacity in the inelastic channel. The latter processes also dominate neutrino heating and cooling, they determine the success or failure of neutrino-driven explosions in multi-dimensional simulations.

\subsection{Supernova EOS}
\label{eos}

The EOS in supernova simulations has to handle a variety of conditions. At low densities and temperatures, time-dependent nuclear burning processes determine the nuclear composition, for which a $\alpha$-network is applied of 20 nuclear species up to $^{56}$Ni including some neutron-rich iron-group nuclei. It is sufficient for an accurate energy generation. Above $T\sim0.5$~MeV nuclear statistical equilibrium (NSE) is achieved and the EOS depends only on the three independent variables temperature $T$, density $\rho$ (or equivalently the baryon density $n_B$), and electron fraction $Y_e$. At low densities, the nuclear composition matches the ideal gas of  $^{56}$Fe or $^{56}$Ni, depending on $Y_e$.  With increasing density and temperatures bulk nuclear matter is reached composed of free nucleons and light nuclear species, in particular $^4$He. The transition to uniform nuclear matter near $\rho_0$ is usually modeled via a phase transition within the nuclear EOS intrinsically.

AGILE-BOLTZTRAN has a flexible EOS module that can handle many currently available baryon EOS. Here I select the relativistic mean-filed (RMF) EOS from Ref.~\cite{Hempel:2009mc} with the RMF parametrization DD2~\cite{Typel:2005,Typel:2009sy}, henceforth denoted as HS(DD2). In addition to the RMF part HS(DD2) is based on the modified NSE for nuclei including the detailed nuclear composition for several thousand species. A comparison with other nuclear statistical models can be found in Ref.~\cite{Buyukcizmeci:2013}. The HS EOS have been explored in supernova simulations in spherical symmetry for a variety of RMF parameterizations~\cite{Hempel:2012,Steiner:2013,Fischer:2014}. In addition to the baryons, contributions from $e^\pm$ and photons are added~\cite{Timmes:1999}.

\begin{figure}
\resizebox{0.50\textwidth}{!}{%
\includegraphics{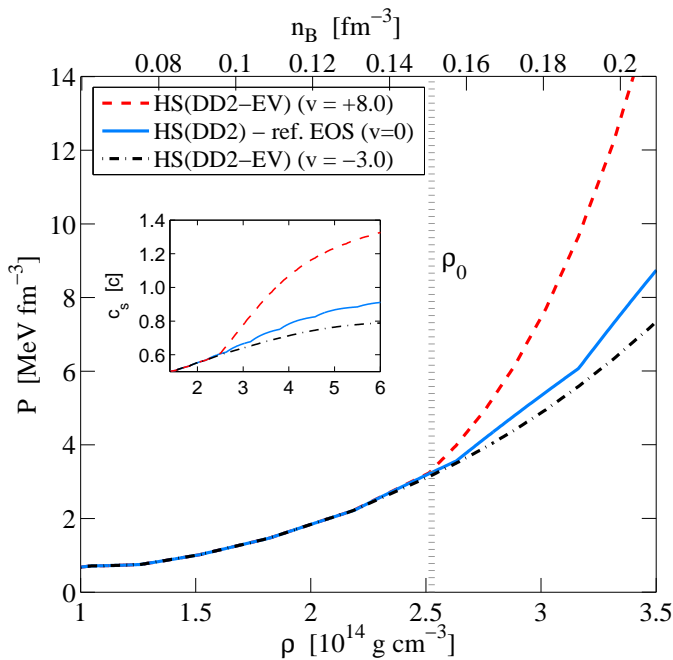}
}
\caption{(color online) High-density supernova EOS under consideration at selected conditions ($T=5$~MeV, $Y_e=0.3$). The vertical dotted line marks saturation density above which the excluded volume modification is active.}
\label{fig:eos}
\end{figure}

\section{Excluded volume extension of the nuclear EOS at finite $T$ and $Y_e$}
\label{ev}

In order to study the super-saturation density EOS in supernova simulations systematically the standard DD2 EOS is extended by taking into account the composite nature of the nucleon. This is hardly possible at the level of the actual degrees of freedom, quarks and gluons. Nevertheless, it can be modeled via an excluded-volume mechanism as discussed in Ref.~\cite{Rischke:1991ke} in the context of RMF models. Usually it is based on a linear functional of the following form, $\phi_i=1-\sum_j \text{v}_j n_j$. It depends on the particle densities $n_j$ and the volume parameter $\text{v}_j$, such that the available volume for the particle species $i$ reduces as follows, $V_i=V\phi_i$, with $V$ being the total volume of the system.

Here, in order to guarantee for a smooth transition the Gaussian type functional of Ref.~\cite{Typel:2015} is employed,
\begin{equation}
\label{eq:phi}
\phi(n_B; \text{v}) =
\left\{
\begin{array}{l l}
1 & \;\;\;\;(n_B\leq \rho_0) \\
\exp\left\{-\frac{\text{v}\vert \text{v} \vert}{2}\left(n_B-\rho_0\right)^2\right\} & \;\;\;\;(n_B>\rho_0)
\end{array}
\right.
\end{equation}
where $\phi_n=\phi_p=\phi$ is assumed. The only parameter is the effective excluded volume v. This formalism is applied for densities in excess of nuclear saturation density, i.e. the sub-saturation density EOS DD2 remains unmodified being in excellent agreement with current experimental constraints (c.f. Ref.~\cite{Lattimer:2013}). The modified volume available for the nucleons generally affects their particle densities ($n_n$, $n_p$) and consequently also their pressure $(p_n,p_p)$ and all other EOS quantities. Further details are given in Ref.~\cite{Typel:2015}.

Furthermore meson and lepton contributions have to be added, both of which are not affected from the excluded nucleon volume modification. The other nuclear EOS quantities, e.g., the energy per baryon and entropy are modified accordingly in order to ensure thermodynamic consistency as well as being still consistent with saturation properties of nuclear matter. It results in a smooth transition from the reference EOS DD2 to the DD2-EV EOS for $\rho>\rho_0$ as illustrated in Fig.~\ref{fig:eos}.  Here we have the flexibility of choosing the excluded volume parameter arbitrarily. It results in stiff and soft EOS with higher and lower pressures at supersaturation densities for $\text{v}>0$ and $\text{v}<0$, compared to the reference case (v=0). I select the parameters v=+8.0~fm$^{3}$ (red dashed line) and \text{v=--3.0~fm$^{3}$} (black dash-dotted line) as two extreme cases, relative to the reference one (blue solid line), as illustrated in Fig.~\ref{fig:eos}. The EOS with excluded volume modifications are henceforth denoted as DD2-EV. Note that the nuclear saturation properties remain unmodified, e.g., with $\rho_0=0.149$~fm$^{-3}$ and symmetry energy $J=31.67$~MeV. However, quantities which relate to derivatives are indeed modified, e.g., the (in)compressibility modulus varies from $K\simeq 541$~MeV (v=+8.0~fm$^3$) to $K\simeq 201$~MeV (\text{v=--3.0~fm$^3$}) compared to the reference case $K\simeq 243$~MeV (v=0).

Both DD2-EV versions explored here reach maximum neutron star masses in agreement with the currently largest and most precise observational pulsar mass constraints of PSR~J1614-2230 ($1.97\pm0.04$~M$_\odot$)~\cite{Demorest:2010} and PSR~J0348-043 ($2.04\pm0.04$~M$_\odot$)~\cite{Antoniadis:2013}. Let me remark here that the parameter v=+8.0~fm$^{3}$ represents indeed the upper limit in terms of stiffness, since the speed of sound ($c_s$) exceeds the speed of light ($c$) above some densities (for details see Fig.~\ref{fig:eos}). The presence of superluminal speed of sound in this model should not become problematic since such high densities are not obtained during the supernova simulations considered here, as will be shown in the following section~\ref{SNsimulations}.

This generalized excluded volume approach affects both the symmetric and asymmetric parts of the EOS above saturation density. Quantities which relate to the symmetry energy are particularly important for weak interactions and the neutrino transport in supernova simulations. However the neutron and proton single particle energies, and in particular their difference, are affected from the excluded volume only mildly as compared to the reference EOS HS(DD2). The same holds for the charged chemical potential, i.e. the difference of neutron and proton chemical potentials. Note further that the here employed elastic approximation for the expressions of charged-current weak interactions (reactions (1) and (2) in Table~\ref{tab:nu-reactions}) use only these quantities, i.e. difference of the neutron-to-proton single particle potentials as well as the charged chemical potential~\cite{Reddy:1998,MartinezPinedo:2012,Roberts:2012}. Hence we cannot expect any impact from the inclusion of the excluded volume on the weak equilibrium obtained at high densities where the neutrinos are trapped. It is determined from the competition of reactions~(1) and (2) in Table~\ref{tab:nu-reactions}. Towards low densities where neutrinos decouple the excluded volume is inactive. I will return to this point when analyzing results from core-collapse supernova simulations in sec.~\ref{SNsimulations}.

\section{Simulation results of the early post-bounce evolution}
\label{SNsimulations}

\begin{figure}
\resizebox{0.49\textwidth}{!}{%
  \includegraphics{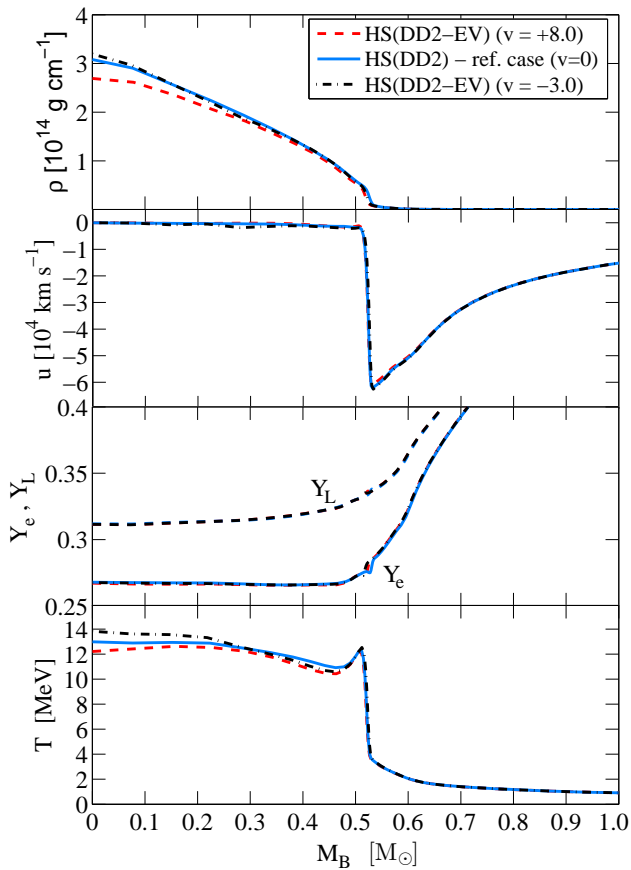}
}
\caption{(color online) Radial profiles of selected quantities at core bounce with respect to the enclosed baryon mass, comparing the different EOS under investigation.}
\label{fig:bounce}
\end{figure}

In the following paragraphs core-collapse supernova simulation results will be analyzed. Focus is on the early post-bounce evolution, i.e. prior to the possible onset of shock revival and subsequent explosion. The simulations start from the 18~M$_\odot$ pre-collapse progenitor model form Ref.~\cite{Woosley:2002zz}. It was evolved consistently through core collapse, bounce and post bounce phases using AGILE -- BOLTZTRAN. I apply the above introduced EOS with the extremely stiff and soft high-density behaviors, HS(DD2-EV) with v=+8.0~fm$^{3}$ (red dashed lines) and v=--3.0~fm$^{3}$ (black dash-dotted lines) respectively, as well as the reference EOS HS(DD2) for which v=0 (blue solid lines). In the following text and figures the units for the excluded volume parameter v will be skipped for simplicity.

In Fig.~\ref{fig:bounce} the first-order impact of the high-density EOS on the dynamics of the collapsing stellar core can be identified. For the stiff EOS (v=+8.0) lowest core densities and temperatures are obtained, compared to the reference case, while the soft EOS (v=--3.0) reaches higher core densities and temperatures. Note the shock position in the velocity profiles in Fig.~\ref{fig:bounce}, separating high-density and low-density domain of the central PNS. The high-density differences have only little consequences for the core electron fraction and lepton fraction $Y_e$ and $Y_L$ respectively. $Y_L$ is determined at the moment when neutrinos become trapped, mainly via neutrino scattering on heavy nuclei. Since the same weak rates were used for all simulations and since the low-density part is identical for all EOS under investigation, the core lepton fraction is expected to remain unaffected (see Fig.~\ref{fig:bounce}). The further decrease of $Y_e$ beyond neutrino trapping is determined from the nuclear free symmetry energy (for a recent review c.f.~\cite{Fischer:2014}). Therefore, the excluded volume modifications of the symmetry energy and associated EOS quantities at super-saturation density are small. This includes, e.g., the neutron-to-proton single particle potential difference and the charged chemical potential. Both of which enter the rate expressions used for the weak reactions (1) and (2) in Table~\ref{tab:nu-reactions} which determine the evolution of the core electron fraction $Y_e$. Hence also the evolution towards weak equilibrium remains unmodified as compared to the reference EOS, which is shown via $Y_e$ at core bounce in Fig.~\ref{fig:bounce}. Note the tiny differences of the shock position in Fig.~\ref{fig:bounce} which are due to a slight mismatch in determining the core bounce.

\begin{figure}
\resizebox{0.49\textwidth}{!}{%
  \includegraphics{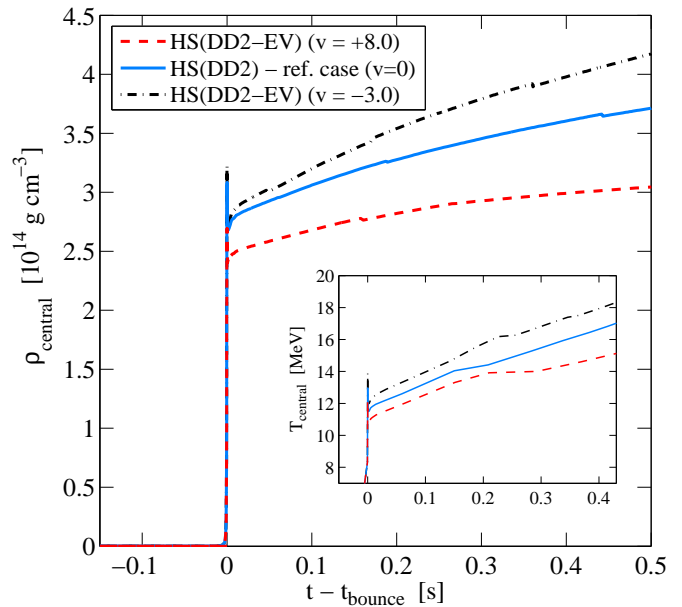}
}
\caption{(color online) Post-bounce evolution of central density and temperature.}
\label{fig:central}
\end{figure}
\begin{table}[htp]
\centering
\caption{Central density and temperature at selected times.}
\begin{tabular}{cccc}
\hline
\hline
$t-t_\text{bounce}$ & v & $T_C$ & $\rho_C$ \\
$[$s$]$ & $[$fm$^3]$ & $[$MeV$]$ & $[10^{14}$ g cm$^{-3}]$ \\
\hline
0 & +8.0 & 12.2 & 2.69 \\
& 0 & 13.0 & 3.08 \\
& -3.0 & 14.0 & 3.22 \\
\hline
0.5 & +8.0 & 15.8 & 3.05 \\
& 0 & 17.0 & 3.71 \\
& -3.0 & 18.7 & 4.17 \\
\hline
\hline
\end{tabular}
\label{tab:SN}
\end{table}

The post-bounce evolution of central density and temperature is illustrated in Fig.~\ref{fig:central} for all EOS under investigation, and in Table~\ref{tab:SN} they are listed at selected times for better comparison. Note that differences between stiff (v=+8.0) and soft (v=--3.0) EOS obtained at core bounce remain also during the post-bounce phase, i.e. with significantly lower and higher core densities, respectively, compared to the reference case. Note in particular the slow-down of the central density rise for the extremely stiff EOS (v=+8.0) which is due to the very steep slope of the pressure gradient for densities in excess of $\rho_0$. Note that unlike inside neutron stars extremely high densities -- several times $\rho_0$ -- are generally not obtained during the early post bounce phase of core-collapse supernovae. Even for the very soft EOS with v=--3.0 the central density reaches only $4.17\times 10^{14}$~g~cm$^{-3}~(1.7\times \rho_0)$ at $t-t_\text{bounce}>0.5$~s. Moreover, the central temperature shows only a marginal response to the excluded volume modification of the high-density EOS. Temperature differences on the order of 1--3~MeV are obtained.

\begin{figure}
\resizebox{0.50\textwidth}{!}{%
  \includegraphics{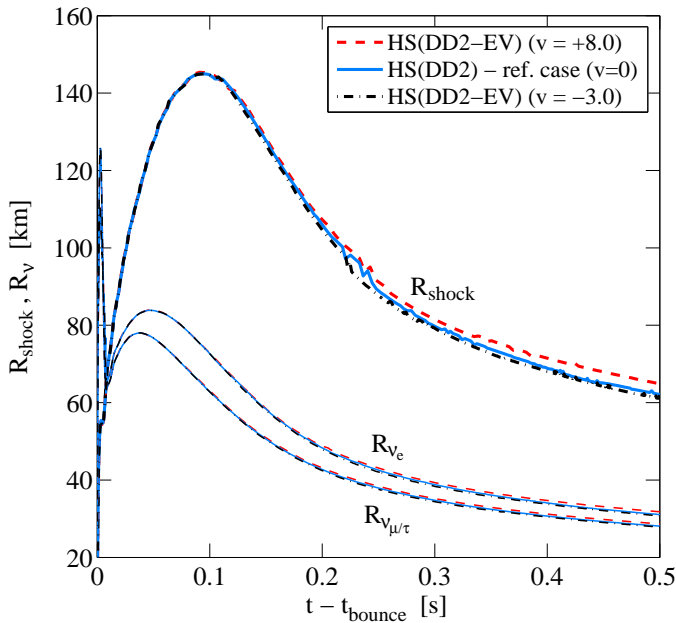}
}
\caption{(color online) Post-bounce evolution of shock radii and neutrinospheres.}
\label{fig:shock}
\end{figure}

\begin{figure}
\resizebox{0.50\textwidth}{!}{%
  \includegraphics{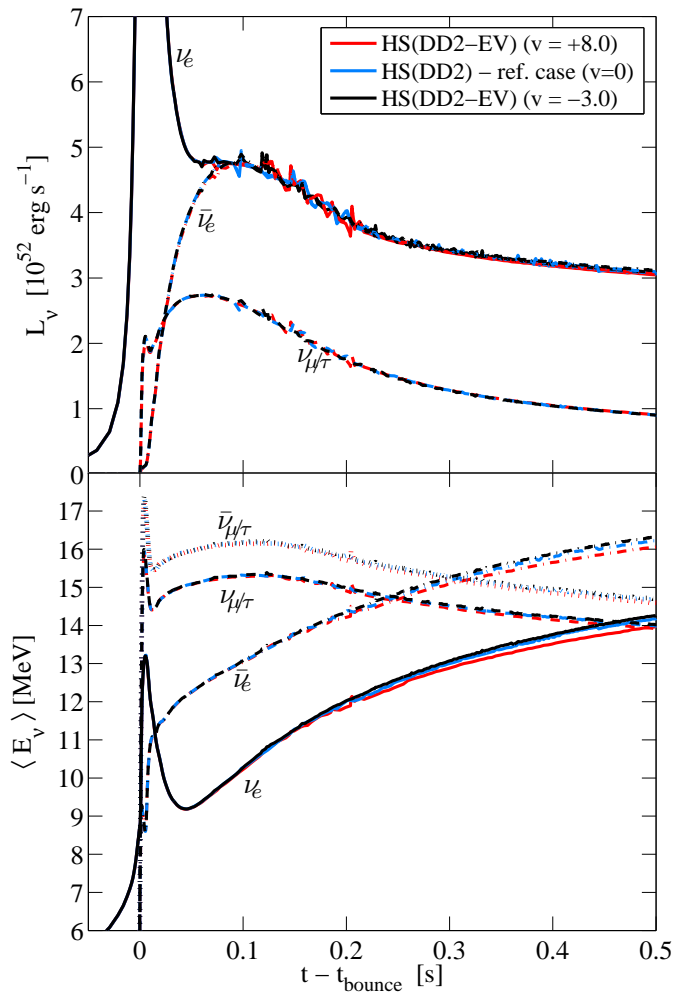}
}
\caption{(color online) Post-bounce evolution of neutrino luminosities and average energies, sampled in the co-moving frame of reference at 500~km.}
\label{fig:neutrinos}
\end{figure}

Despite the large differences of the central density obtained for the different excluded volume parametrization during the post-bounce simulations, it has only little impact on the PNS structure. Enclosed mass as well as shock positions and neutrinosphere radii are only mildly affected towards later times, $t-t_\text{bounce}>0.3$~s (see Fig.~\ref{fig:shock}). The relevant physics of core-collapse supernovae takes place at sub-saturation density, i.e. where the evolution of PNS contractions and supernova shock dynamics is determined from neutrino heating and cooling. In particular, the contraction behavior of the PNS is driven by the accretion of low-density material onto its surface, from the gravitationally unstable layers above the stellar core. It also defines the neutrino luminosities and spectra of $\nu_e$ and $\bar\nu_e$ which decouple inside this layer of low-density accumulated material at the PNS surface. Neutrinos trapped at higher densities inside the PNS interior cannot diffuse out on timescale on the order of 100~ms. Hence, the contraction of the high-density part of the PNS cannot affect significantly the supernova dynamics nor the neutrino signal (see therefore Fig.~\ref{fig:neutrinos}).

Only after $t-t_\text{bounce}>0.3$~s the high-density PNS contraction starts to affect the low-density envelope, mainly due to a somewhat stronger(weaker) gravitational potential for the soft(stiff) EOS which reach higher(lower) central densities. This leads to a slightly faster(slower) shock withdraw (see Fig.~\ref{fig:shock}). Differences can be also identified on the order of less than100~keV lower(higher) average neutrino energies for the electron flavors for $v=+8.0$($v=-3.0$) compared to the reference simulation. Heavy lepton flavor neutrinos, which decouple at generally higher densities, are consequently less affected (see Fig.~\ref{fig:neutrinos}).

\section{Summary and conclusions}
\label{summary}

In this article the impact of the high density EOS on the dynamics of core-collapse supernova simulations as well as on the potentially observable neutrino signal is studied. Standard nuclear EOS with hadrons and mesons as degrees of freedom are based on the point-like quasi-particle picture, e.g., within the RMF framework. Such nuclear model DD2 was employed here as the reference case. Extending the simple quasi-particle picture by considering the composite nature of the baryons is not feasible at the level of quark and gluon degrees of freedom, especially under supernova conditions. In this study their impact is approximated via the novel generalized excluded volume approach of Ref.~\cite{Typel:2015}. It results in modifications of the well calibrated RMF EOS DD2 at supersaturation densities. The excluded volume supernova EOS versions, HS(DD2-EV), depend only on the excluded volume parameter. Here I select two variations which result in a extremely stiff and in another extremely soft EOS at supersaturation density. However, they are adjusted to be still in agreement with nuclear constraints, e.g., nuclear matter properties at $\rho_0$ as well as with observations of $\sim2$~M$_\odot$ neutron stars.

The EOS HS(DD2-EV) are explored in simulations of failed core-collapse supernova explosions during the early post-bounce phase. This phase is determined by mass accretion onto the central PNS, where a thick layer of low-density material accumulates at the PNS surface. The latter contracts accordingly on a timescale on the order of several 100~ms. Unlike initial expectations this study confirms that the high-density domain of the PNS has a negligible impact on the PNS contraction behavior. Despite large differences at supersaturation density the supernova evolution in terms of shock dynamics as well as the neutrino luminosities and energies are affected only marginally and in particular only towards late times. Previous studies of the nuclear EOS role in supernova simulations were based on models which differ in many (if not all) nuclear matter properties. This made it difficult to identify the high-density EOS impact on potential supernova observables. With this novel excluded volume approach only the supersaturation density EOS is affected and in particular the low density EOS of HS(DD2-EV) remain unmodified. For the first time this allows for the direct identification of the supersaturation density EOS influence, despite the non-linearity of hydrodynamics and neutrino transport. In addition to the 18~M$_\odot$ intermediate-mass progenitor discussed above, a low mass progenitor of 11.2~M$_\odot$ was considered for the same EOS HS(DD2-EV). For this one differences of the PNS evolution are even smaller, mainly because central densities are generally somewhat lower.

Note that qualitatively similar conclusions were obtained in previous studies~\cite{Swesty:1994,Thompson:2003} which were based on the commonly used supernova EOS from Ref.~\cite{Lattimer:1991nc}. It is available to the community for three different values of the (in)compressibility modulus, $K=180/220/375$~MeV. In this sense they explore the stiffness of the EOS, however, also at subsaturation density. The conclusions drawn form the present analysis are due to significantly larger variations of the (in)compressibility modulus ($K=201-541$~MeV). Note that the very soft version with $K=180$~MeV is violating several constraints, e.g., the maximum neutron star mass is too low and it is in large disagreement with the neutron matter EOS constraint from chiral EFT~\cite{Tews:2013,Krueger:2013}. The latter constraint is also violated for the version with $K=220$~MeV, in particular at low densities relevant for the supernova dynamics (c.f. Fig.~3 of Ref.~\cite{Fischer:2014}).

Focus of this study relates to conditions where to first order multi-dimensional phenomena can be neglected. Their main contribution is at the low density regime, in terms of turbulent hydrodynamics, where the EOS remains unmodified due to the excluded volume. The here presented conclusions are unlikely to change when the multi-dimensional nature of hydrodynamics and neutrino transport is taken into account. Even though the magnitude of here presented observables may well be altered, relative changes as well as conclusions are expected to remain qualitatively.

The central PNS densities reached during a canonical core-collapse supernova post bounce mass accretion phase are significantly lower than those of cold neutron stars. The difference is due to temperatures in excess of 10~MeV and the large component of trapped neutrinos of all flavors. This, in combination with the here presented analysis, leads to the conclusion that core-collapse supernova studies can be excluded as laboratories to efficiently probe the supersaturation density state of matter for EOS that remain continuous towards higher densities. Alternatively the presence of a discontinuety, e.g., via a (strong) 1st-order phase transition at supersaturation density perhaps to deconfined quark matter may be identified via the neutrino signal~\cite{Sagert:2008ka,Fischer:2011} and/or complementary via the gravitational wave signal.

\section*{Acknowledgement}

Special thanks belongs to S.~Typel for providing the EOS tables. The supernova simulations were performed at the LOEWE-Center for Scientific Computing in Frankfurt, Germany. The author also acknowledges support from the Polish National Science Center (NCN) under grant number UMO-2013/11/D/ST2/02645.


\end{document}